# An Efficient Algorithm for Automatic Structure Optimization in X-ray Standing-Wave Experiments

Osman Karslıoğlu[1], Mathias Gehlmann[2,3], Juliane Müller[4], Slavomir Nemšák[5], James A. Sethian[6,7], Ajith Kaduwela[8], Hendrik Bluhm[1,5], Charles Fadley*[,2,3]

[1]Chemical Sciences Division, Lawrence Berkeley National Laboratory, Berkeley CA, USA
[2]Material Sciences Division, Lawrence Berkeley National Laboratory, Berkeley CA, USA
[3]Department of Physics, University of California Davis, Davis CA, USA
[4]Computational Research Division, Lawrence Berkeley National Laboratory, Berkeley CA, USA
[5]Advanced Light Source, Lawrence Berkeley National Laboratory, Berkeley CA, USA
[6]Department of Mathematics, University of California Berkeley, Berkeley CA, USA
[7]Center for Applied Mathematics for Energy Research Applications,
Lawrence Berkeley National Laboratory, Berkeley CA, USA
[8]Air Quality Research Center, University of California Davis, Davis, CA USA
* Corresponding author (fadley@physics.ucdavis.edu)

## Abstract

X-ray standing-wave photoemission experiments involving multilayered samples are emerging as unique probes of the buried interfaces that are ubiquitous in current device and materials research. Such data require for their analysis a structure optimization process comparing experiment to theory that is not straightforward. In this work, we present a new computer program for optimizing the analysis of standing-wave data, called SWOPT, that automates this trial-and-error optimization process. The program includes an algorithm that has been developed for computationally expensive problems: so-called black-box simulation optimizations. It also includes a more efficient version of the Yang X-ray Optics Program (YXRO) [Yang, S.-H., Gray, A.X., Kaiser, A.M., Mun, B.S., Sell, B.C., Kortright, J.B., Fadley, C.S., J. Appl. Phys. 113, 1 (2013)] which is about an order of magnitude faster than the original version. Human interaction is not required during optimization. We tested our optimization algorithm on real and hypothetical problems and show that it finds better solutions significantly faster than a random search approach. The total optimization time ranges, depending on the sample structure, from minutes to a few hours on a modern laptop computer, and can be up to 100x faster than a corresponding manual optimization. These speeds make the SWOPT program a valuable tool for realtime analyses of data during synchrotron experiments.

# 1. Introduction

X-ray standing-wave techniques have been developed in the 1960s, and since then blossomed into a powerful set of tools for investigating the electronic structure of materials with high spatial resolution. Initially, Bragg reflections from crystal lattices were used to selectively probe the structure of materials on the atomic scale, using X-ray fluorescence or Auger emission as the probe [1-4]. Later, periodic polycrystalline samples — such as magnetic multi-layered materials — were used to investigate structural properties with unprecedented depth selectivity, using reflectivity, fluorescence, and X-ray absorption as probes [5].

The use of photoemission (PES) in standing-wave studies came somewhat later beginning in the 1990s with single-crystal systems and hard/tender X-rays with energies of about 3 keV [6]. This method was then extended into multilayer systems with soft X-rays [7]. The small inelastic mean free path of photoelectrons — as generated by soft X-rays — renders the PES technique very surface sensitive. Even though surface sensitivity is undesirable in experiments probing bulk or buried-interface properties, the PES sensitivity to chemical states, electric fields and electron momentum makes it an irreplaceable tool for studying core-level and valence band properties and enables direct mapping of the electronic band structure. For example, there are by now approximately 40 papers on soft- and hard- X-ray SW-PES in the literature, including both single-crystals and synthetic heterostructures, as reviewed elsewhere [8, 9].

One of the latest expansions of SW-PES was angle-resolved studies of valence states. The technique successfully addressed emergent phenomena at complex oxide interfaces such as LSMO/STO [10], GTO/STO [11], and LNO/STO [12]. Another important development was combining standing-waves with ambient-pressure XPS to study ion adsorption [13] and electro-oxidation [14] at solid/liquid interfaces. More recently, standing-wave hard X-ray photoemission was used to obtain site- and element-specific band structure in single crystals of a dilute magnetic semiconductor: Mn-doped GaAs [15].

The experimental data in SW-PES are usually collected in the form of core-level photoelectron intensities as a function of incidence angle (aka rocking curves or RCs) or photon energy. First-order Bragg reflection creates a strong periodic modulation in the exciting electric field, and by slightly changing the incident angle or the photon energy one is able to scan the position of the standing-wave by half of its period. Spatially resolved information is obtained by correlating the experimental data with calculations using a specially written X-ray optical program called YXRO [16].

Using multi-layer mirrors as the SW generators — with layers of interest grown within or transferred onto them — brings certain challenges. Apart from uncertainties in the structural parameters of the sample itself, one has to consider the imperfections present in the mirror. However, using photoemission rather than simple reflection/diffraction/absorption data brings further dimensions of information to the experiment; each chemical species exhibits a distinctive rocking curve, as they originate from different depths and their locations with respect to the nodes/antinodes of the standing-wave are generally different. For example, recent work that combined X-ray standing-wave photoemission with ambient-pressure PES (SWAPPS) resulted in unprecedented accuracy in probing solid/liquid interfaces [13]; this demonstrates the ability of the SWAPPS technique to study more complicated systems that involve chemical species that are free to move in a liquid layer, in contrast to the fixed layers in a solid.

The task of obtaining the theoretical RCs that best match the experimental RCs is essentially an optimization problem with the goal to minimize the discrepancy between theory and experiment by determining the individual layer thicknesses and upper and lower interdiffusion lengths/roughnesses. An analytical measure of this discrepancy and its derivatives are not available. Thus, a derivative-free global optimization method is needed to avoid falling into a false minimum in the search. The parameter space is proportional to the complexity of the problem at hand. The state of the art approach to solving these problems is to use chemical intuition and sheer computational force. But this can engender human error in searching the parameter space, and also can be very time consuming. In this work, we present an efficient, automated method to address this problem. We use a global optimization algorithm that was developed for computationally expensive black-box simulation optimization problems, and for which analytical expressions of the objective function and its derivatives are not available. The result is a computer program for optimizing the analysis of SW data called SWOPT that accepts a problem definition — including the experimental RCs and known estimates of the structure of the system — and returns the best solution without any intermittent user input. SWOPT uses YXRO's original computation code [16] without the graphical user interface (GUI), a black-box optimizer (BBO), and a custom interface that orchestrates the data transfer between the user, YXRO, and the BBO.

To illustrate the capabilities of SWOPT, we have used it to solve three example problems and we show a clearly better performance in each case, as compared to a random search method, which uses sheer computational force. The increased speed of YXRO and the elimination of user interaction makes SWOPT very attractive for realtime data analyses during SW-PES synchrotron experiments.

## 2. Computational Approach

In the following, we describe the general structure of SWOPT. Figure 1 shows a flowchart of the program, which is an integrated MATLAB® code that consists of YXRO, the BBO and an interface between the two and the user. We provide more details in the following subsections.

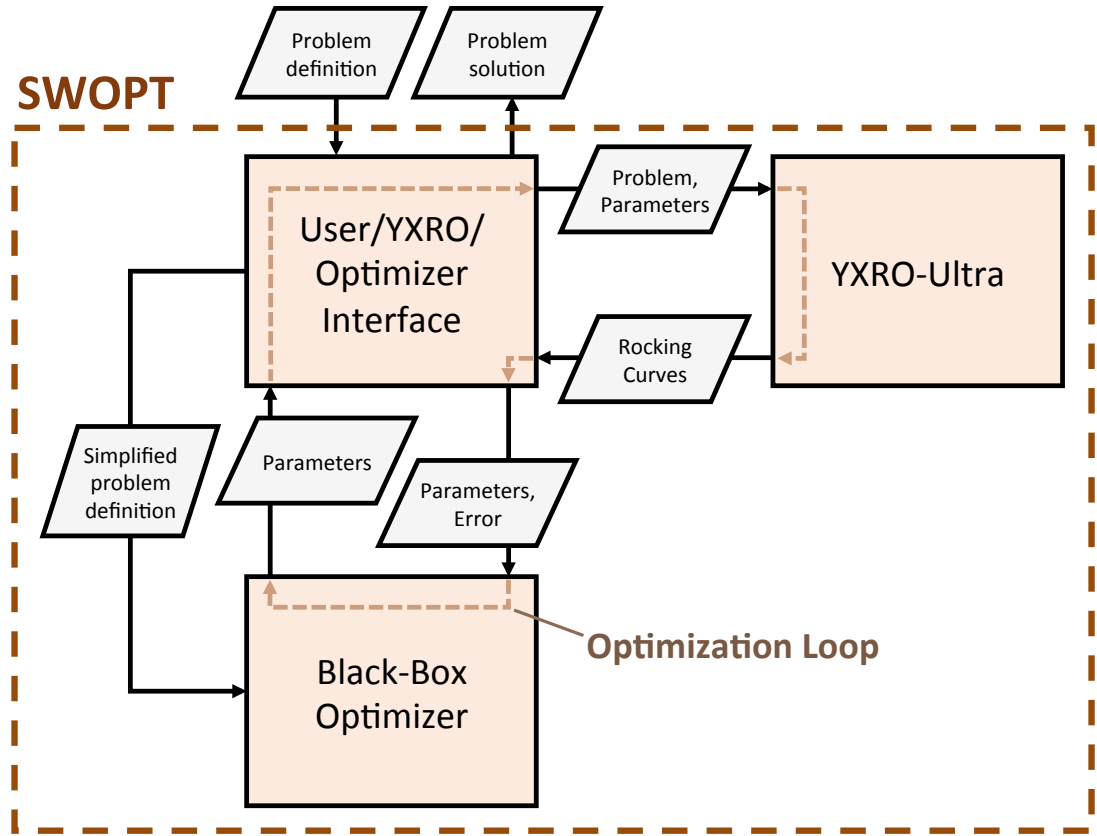

Figure 1: The general structure of the computer program SWOPT, which solves structure optimization problems using YXRO and a black-box optimizer.

### 2.1. YXRO-Ultra

YXRO was originally created as a standalone MATLAB® program that is used through a GUI [16]. For problems where hundreds or even thousands of simulations need to be run to find the optimum structure, the initiation and termination of the GUI adds considerable computation time which is not spent on simulating RCs. Furthermore, YXRO does not provide the kind of flexibility that is needed for complex problems like solid/liquid interfaces, where solution phases overlap one another in depth and need to have their distributions optimized. To address these issues, we used a modified source code of YXRO — which is still written in MATLAB® — that runs without the GUI and is completely accessible to a higher-level optimization code. We call the resulting program "YXRO-Ultra". YXRO-Ultra is computationally about an order of magnitude more efficient than YXRO, not only due to the removal of the GUI, but probably also due to the elimination of the old, separate runtime system.

Below, we provide the details of the operation of the program.

## 2.2. User Input/Problem Definition

The problem definition is provided by the user and is based on a so-called template file. This template is a regular YXRO input file, which can be generated either with the graphical user interface of YXRO or within the YXRO-Ultra scripts. The template defines basic sample properties such as layers with their composition and optical properties, and it defines the experimental conditions such as electron binding energies, photon energy, angles of X-ray incidence and electron exit, etc.

The next step for the user is to define the optimization parameter ranges, which will be varied by the BBO to modify the sample properties or experimental conditions based on the template. The user sets the range in which each parameter should be varied, and since these physical properties cannot be measured beyond a certain accuracy, the user also defines the step size for sampling each parameter. In this way we discretize the search space, reducing the total number of possible solutions to a finite set. In most cases these parameters are layer thicknesses and interdiffusion lengths within the sample. In addition to the range of individual parameters, the problem definition allows for constraints. These are analytical conditions that will render a parameter vector invalid if they are violated. An example would be the relationship between the elemental and native oxide layer thicknesses for a Si capping layer. That is, when elemental Si at the top of a Si layer oxidizes, the volume per Si atom increases by a factor of 2.20 due to the incorporation of O atoms into the structure. This can be translated into analytical terms as $\Delta t_{SiO2} = -2.20 \Delta t_{Si}$, where $\Delta t$ denotes the layer thickness change of both layers, relative to the situation with no oxide.

For most problems, such as examples 1 and 2 in this article, the definition of thickness and interdiffusion/roughness parameters in combination with constraints are sufficient to determine the sample properties. However, the script-based problem definition, which directly uses YXRO-Ultra, allows the user to include and modify any property of the sample or the experiment that is defined within the YXRO program. Therefore, the user can also set up more advanced properties as optimization parameters, such as thickness gradients, material densities, or electron inelastic mean free paths. In example 3 we make use of this by optimizing the distribution of two independent chemical elements within a water layer.

## 2.3. User/YXRO/Optimizer Interface

This interface connects the user input with the BBO and YXRO-Ultra in the following way:

1. It interprets the problem definition of the user and creates a simplified definition for the black-box optimizer.

2. It receives parameters from the optimizer and translates them into a YXRO input.
3. It receives the YXRO output (i.e. rocking curves or energy scans), calculates the error (i.e. the objective function for the optimizer), and returns this error to the optimizer.
4. For each trial, it returns the problem solution to the user, as plots and data files.
5. Guided by the optimizer, it loops on parameters that lie within the ranges specified by the user, until a desired degree of fit is obtained, as judged by a sum of R factors

The interface interprets the broad problem definition and presents the optimizer a simplified problem. The mathematical formulation of the optimization problem from the optimizer's perspective is as follows:

$$\min_x f(x) \tag{1}$$

$$c_j(x) \leq 0, j = 1, \ldots, m \tag{2}$$

$$x_i \in \{x_i^l, \ldots, x_i^u\}, i = 1, \ldots, d, x_i \in \mathbb{N} \tag{3}$$

where $f(x)$ is the overall error as summed over all of the $d$ optimization parameters, $x_i^l$ and $x_i^u$ are the lower and upper bounds of the $i^{th}$ optimization parameter. These parameters can be layer thicknesses and interdiffusion lengths. The constraints $c_j(x)$ are given analytically and if violated, they disqualify parameter vectors (an example would be the relationship between the elemental and native oxide layer thicknesses for a Si capping layer discussed above). For the black-box optimizer, we transform the search space such that all parameters are integers ($\mathbb{N}$ denotes the set of natural numbers), with a trivial back transformation to report results to the user.

The optimizer takes the simplified problem definition and returns a set of initial parameters to evaluate with YXRO. The interface uses these parameters and the problem definition to generate a YXRO input, sends it to YXRO, and receives the simulated RCs as output. It then compares the YXRO output to the experimental RCs and calculates the error $f(x) = R$ using the following expressions:

$$f(x) = R = \sum_{j=1}^{Q} R_j \tag{4}$$

$$R_j = w_j \sum_{i=1}^{N} \left(I_{j,i}^{\text{Exp}} - I_{j,i}^{\text{Sim}}\right)^2 \tag{5}$$

where $I$ is the signal intensity (typically normalized to be unity outside of the Bragg reflection regime), $N$ is the number of angle or energy points in a RC, and $Q$ is the total number of RCs. $w_j$ is a weighting factor with a default value of 1, but can also be input by the user to give more weight to experimental RCs with higher counts (i.e., less statistical noise). Other formulas for R factors could also in principle be defined by the user in the future, such as that given by Pendry for low energy electron diffraction [17],

An Efficient Algorithm for Automatic Structure Optimization in X-ray Standing-Wave Experiments

$$R_j = \frac{\sum_{i=1}^{N}\left(I_{j,i}^{Exp}-I_{j,i}^{Sim}\right)^2}{\sum_{i=1}^{N}\left(\left(I_{j,i}^{Exp}\right)^2+\left(I_{j,i}^{Sim}\right)^2\right)} \tag{6}$$

Other R-factors and normalizing expressions that are discussed elsewhere might also be used [18].

In deriving the intensities $I_{j,i}^{Exp}$ in Eq. (5), the raw RCs are typically normalized by one of the three available normalization options: by dividing each RC by i) the mean of the RC, ii) a polynomial background which is typically fitted through the non-modulating part of the RC, or iii) the maximum of the RC. These procedures are used to eliminate angle-dependent instrumental effects, e.g. having to do with the extent of the emitting spot on the sample as seen by the spectrometer. For consistency, the same procedure is applied to the raw simulated RCs to obtain $I_{j,i}^{Sim}$.

$R$ is the objective function $f(x)$ that the optimizer minimizes. The interface returns the $R$ value for a given parameter set to the optimizer and receives a new set of parameters from it. YXRO runs a simulation for these parameters, and this cycle continues until the optimizer reaches a satisfying condition, which is currently a maximum number of simulations. At that point, the best parameter vector is returned to the interface and the interface creates a summary of the problem solution for the user.

## 2.4. Black-Box Optimizer

When faced with computationally expensive black-box optimization problems, one widely adopted approach is to use surrogate models to inform the optimization search. A surrogate model $s(x)$ is a computationally cheap approximation of the time-expensive simulation objective function $f(x)$, [19], with $f(x)$ requiring in general many iterations of the YXRO program:

$$f(x) = s(x) + e(x). \tag{7}$$

Here, $e(x)$ is the difference between the time-expensive simulation function and the time-cheap surrogate model. Different types of surrogate models have been developed in the literature, among others radial basis functions [20], kriging [21], multivariate adaptive regression splines [22], polynomial regression models [23], and ensembles of different models [24]. Our surrogate integer optimization algorithm works similar to the SO-I scheme introduced elsewhere [25]. In this paper, we extend the functionalities of SO-I to take analytical fast-to-compute constraints into account.

Generally, any type of surrogate model could be used, but in this work, we use radial basis function (RBF) surrogate models [20], because our objective function is deterministic and RBFs are interpolating and they therefore predict the true function value at an already evaluated point. An RBF surrogate model is defined as follows:

$$s(x) = \sum_{k=1}^{n} \lambda_k \phi \left( ||x_k - x||_2 \right) + p(x), \tag{8}$$

where $x_k, k = 1, \ldots, n$, are the parameter vectors at which we have already evaluated the expensive simulation objective function, $\phi(\cdot)$ is the radial basis function (here, we use the cubic, $\phi(r) = r^3$), and $p(x)$ is the polynomial tail whose order depends on the type of RBF we have chosen (for the cubic RBF, we need at least a linear polynomial, $p(x) = \beta_0 + \beta^T x$). The parameters $\lambda_k, k = 1, \ldots, n, \beta_0,$ and $\beta = [\beta_1, \ldots, \beta_d]^T$ are determined by solving a linear system of equations:

$$\begin{bmatrix} \Phi & P \\ P^T & 0 \end{bmatrix} \begin{bmatrix} \lambda \\ \gamma \end{bmatrix} = \begin{bmatrix} F \\ 0 \end{bmatrix}, \tag{9}$$

where,

$$P = \begin{bmatrix} x_1^T & 1 \\ x_2^T & 1 \\ \vdots \\ x_n^T & 1 \end{bmatrix}, \lambda = \begin{bmatrix} \lambda_1 \\ \vdots \\ \lambda_n \end{bmatrix}, \gamma = \begin{bmatrix} \beta_1 \\ \vdots \\ \beta_d \\ \beta_0 \end{bmatrix}, F = \begin{bmatrix} f(x_1) \\ \vdots \\ f(x_n) \end{bmatrix}, \tag{10}$$

and $\Phi_{\iota,\nu} = \phi(||x_\iota - x_\nu||_2), \iota, \nu = 1, \ldots, n.$ $||\cdot||_2$ denotes the Euclidean norm. The matrix in the linear system in (9) is invertible if and only if $\text{rank}(P) = d + 1$.

We use the RBF models in our iterative optimization algorithm in order to only select the most promising parameter vectors for evaluation (promising in the sense that it is the solution to a computationally cheap auxiliary optimization problem). The optimization algorithm takes into account all constraints $c_j, j = 1, \ldots, m,$ and all integrality constraints. Hence, we do not waste time calculating the objective function for parameter vectors that are clearly unphysical. The steps of the algorithm are illustrated in Figure 2 and the details are summarized in the appendix.

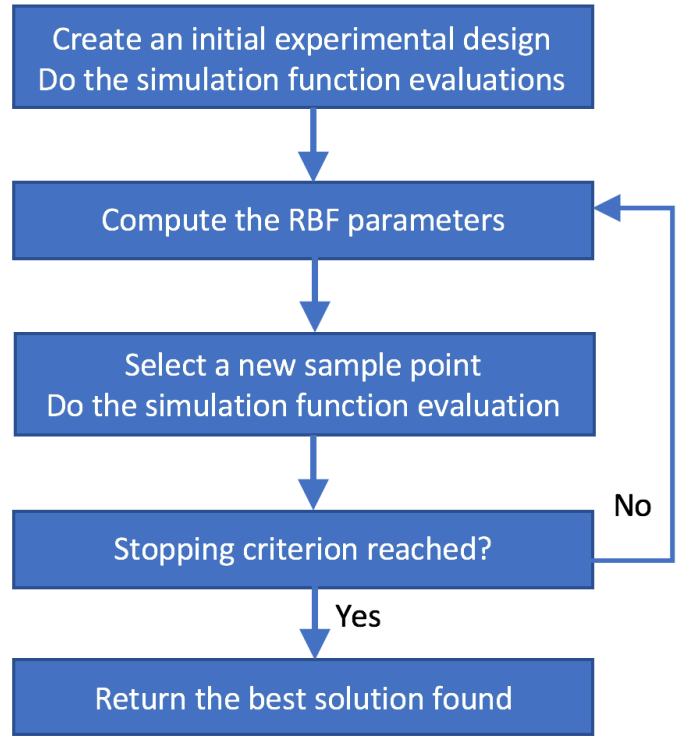

Figure 2: Flowchart of the optimization algorithm (BBO).

A graphical illustration of the individual steps of the BBO for a one-dimensional example is provided in Figure 3 below. Note that this method tends to avoid getting trapped in false local minima. For example, it does not just focus the search on the immediate neighborhood of x ≈ 14.7 in this figure.

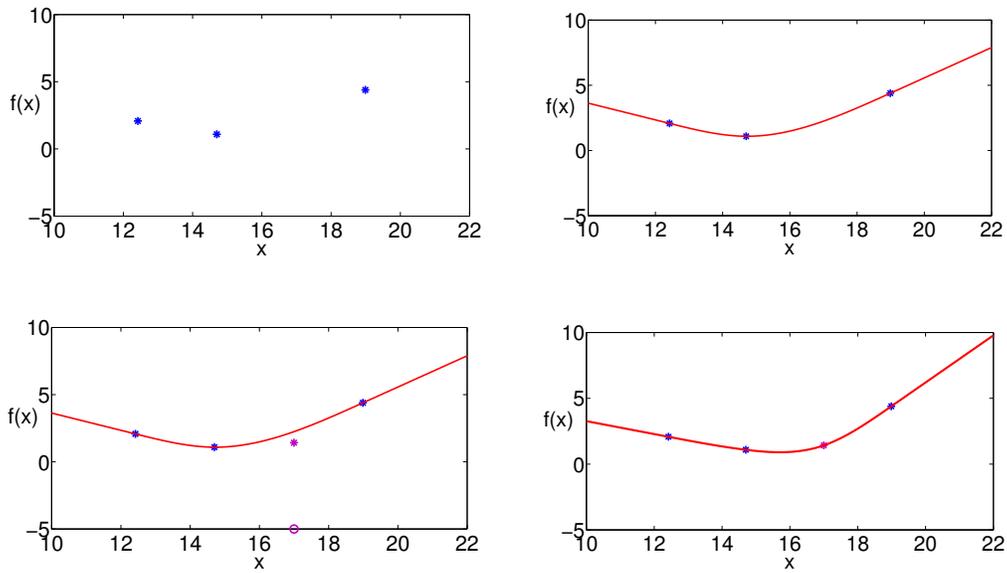

Figure 3: One-dimensional illustration of the surrogate model based optimization search. Initial experimental design (top left); fit a surrogate model (top right); select and evaluate a new point (bottom left); update the surrogate surface with the new information (bottom right).

## 3. Example Applications

In this section, we will demonstrate the effectiveness of SWOPT on three examples. In each example, we will compare the performance of our black-box optimization algorithm to a simple brute-force random search of structural parameter vectors. Our aim is to demonstrate that the adaptive parameter selection of BBO works significantly better than brute-force approaches such as manually varying only a few structural parameters — as done previously for one system [13] — or randomly sampling the structural parameters over the entire space.

The calculations for the following examples were performed using an Intel® CoreI i5-2500 CPU @ 3.30GHz. The parallelization of our program used 4 cores of the CPU for the YXRO simulation. The total calculation time is mostly defined by the YXRO simulations, which depend on the experimental scenario. The average time per evaluation in the following examples ranged from 1.4 s in example 1 to 2.2 s in example 3. In this article we allowed a total number of 10,000 evaluations for each trial, which means that the total optimization time is ~4–6 h each. However, as we will show in the following, in most cases a satisfactory solution can be found with fewer evaluations, and greater parallelization in the future can lead to further reduction in optimization times. For most cases, our program will allow for a realtime analysis of SW data on laptop computers and yields acceptable optimization results within ~1–2 h.

### 3.1. Example 1: An Artificial Test Case — Ideal Sample on Si/Mo Multilayer Mirror

The first example is a hypothetical Si/Mo multilayer mirror (MLM) where we generate "pseudo-experimental" data with YXRO (i.e. we know the optimal solution). We expect that SWOPT will find the exact structure since the data have no noise and can be reproduced exactly by YXRO. However, there may be other near-optimal solutions. This example serves as proof-of-concept to test the accuracy of the optimization method.

The problem definition is given in Figure 4a. The photon energy for this example is in the hard/tender X-ray regime at 3500 eV. The hypothetical sample is 5 layers of metal (from top to bottom: Au, Pt, Ti, Pt, Ti) on top of a Si/Mo multilayer mirror of 80 repeats, where the Mo and Si layer thicknesses are fixed at 9 and 25 Å, respectively. The interfaces of all layers are defined to be sharp. During optimization, each Au, Pt, and Ti layer thickness can vary from 0 to 35 Å, with a step size of 0.5 Å. One complication of this problem is that Pt and Ti signals come from two sets of distinctly separated layers and are integrated to give only one RC for each element. The Pt and Ti RCs depend strongly on how thick each of these layers are, whereas the Au RC depends on the total thickness of Pt+Ti layers. Another complication of the problem is that neither the pseudo-experimental nor the simulated RCs are normalized. Thus, the optimizer has to match the exact amplitudes, rather than relative amplitudes. This is expected to drive the optimizer more to the global minimum rather than to local minima.

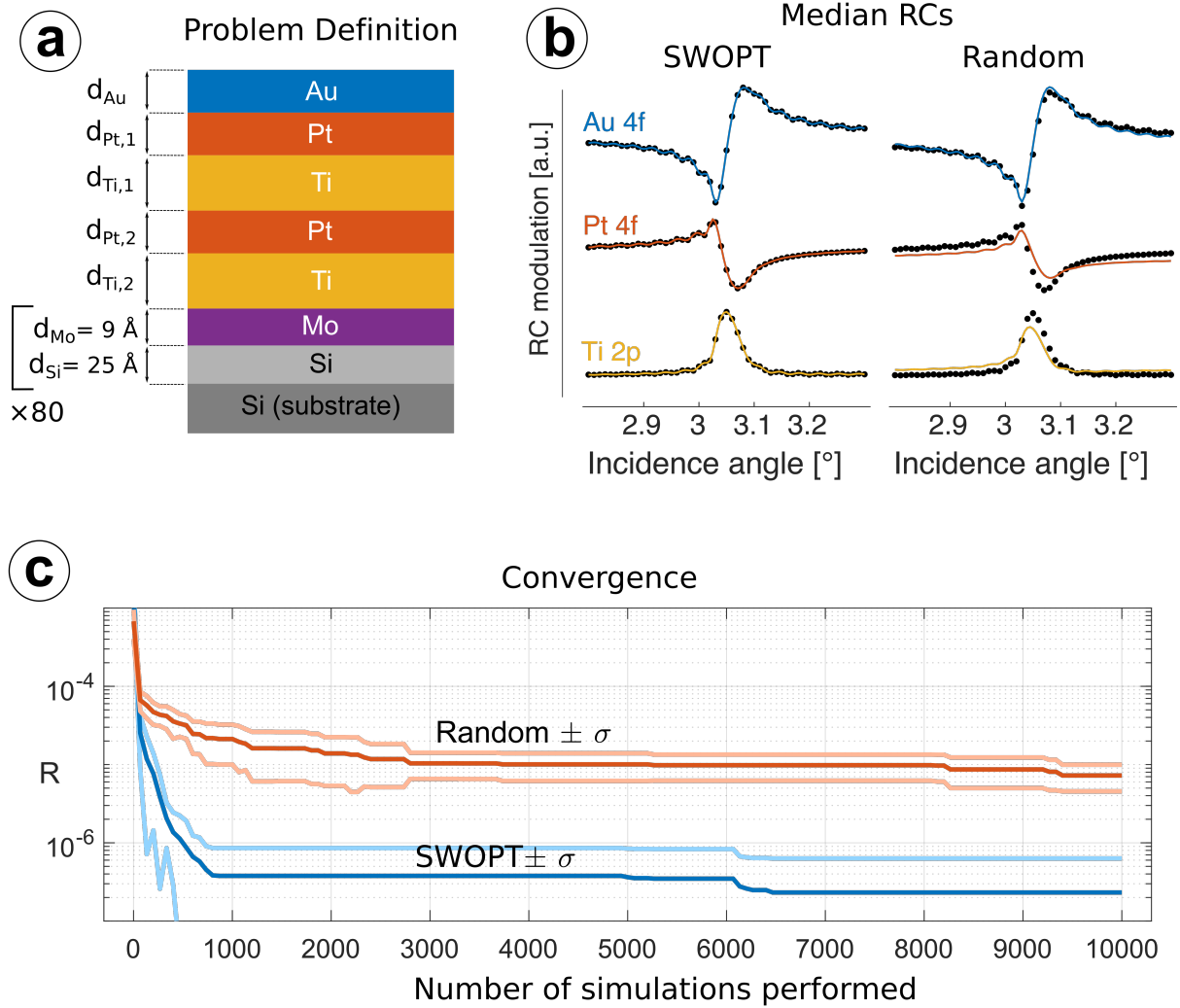

Figure 4: (a) Problem definition for Example 1, (b) pseudo-experimental data (solid circles), SWOPT-found (curves-left), and random-search-found (curves-right) rocking curves for Au 4f, Pt 4f, and Ti2p, (c) best R value found as a function of the number of simulations performed for SWOPT and random search. The darker curves represent the mean R values of 10 different runs of 10,000 simulations each, whereas the lighter curves indicate standard deviation around the means. "$\langle R \rangle - \sigma$" for SWOPT becomes negative after ~500 simulations and thus has been left out of the plot, which is logarithmic.

We ran SWOPT and the random search 10 times each with random initial guesses and with 10,000 YXRO simulations each time. The RC's of the median (5[th] best) of 10 solutions are given in Figure 4b. SWOPT found the exact solution whereas the random search found a solution that almost matches the phase but not the intensities. The convergence behavior of the two algorithms can be seen in Figure 4c. SWOPT approaches the optimal solution significantly faster and with a standard deviation that is significantly smaller (note the logarithmic scale of the y-axis) than the random search method.

Table 1: Layer thicknesses obtained from SWOPT and the random search algorithm ("Random") for the artificial test case. "Median" refers to the thicknesses obtained in the 5$^{th}$ best out of 10 solutions of SWOPT and random. "min/max" refers to the minimum/maximum value found for the parameter within the 10 runs. Mean values are reported for completeness; they are ensemble averages and don't necessarily correspond to a solution of SWOPT or random search.

|  | true value [Å] | median [Å] | | mean (st.dev.) [Å] | | min [Å] / max [Å] | |
|---|---|---|---|---|---|---|---|
|  |  | SWOPT | Random | SWOPT | Random | SWOPT | Random |
| $d_{Au}$ | 11.0 | 11.0 | 12.0 | 11.1 (0.2) | 11.5 (0.5) | 11.0 / 11.5 | 11.0 / 12.0 |
| $d_{Pt,1}$ | 13.0 | 13.0 | 11.0 | 12.9 (0.3) | 12.4 (2.9) | 12.5 / 13.5 | 5.5 / 15.0 |
| $d_{Ti,1}$ | 17.0 | 17.0 | 8.5 | 16.6 (1.0) | 18.9 (7.8) | 14.5 / 17.0 | 8.5 / 29.5 |
| $d_{Pt,2}$ | 7.0 | 7.0 | 5.0 | 7.2 (1.2) | 7.4 (7.1) | 4.5 / 8.5 | 0.0 / 23.5 |
| $d_{Ti,2}$ | 22.0 | 22.0 | 34.5 | 22.4 (1.8) | 20.6 (8.3) | 21.0 / 26.5 | 7.0 / 34.5 |

The values found for the five layer thicknesses defined in the problem are given in Table 1. The median solution found by SWOPT is essentially exact, whereas the one found by the random search algorithm is significantly inaccurate for the bottom three layers. The inadequacy of the random search can also be seen in the standard deviations, which are 40-96% for the bottom three layers. The standard deviations increase from the top to the bottom layers for both SWOPT and random search.

## 3.2. Example 2: Si-Mo Multilayer Mirror

The second example is a real pristine MLM consisting of 80 Si/Mo bilayers. The mirror was prepared at the Center for X-ray Optics at Lawrence Berkeley National Laboratory using magnetron sputtering on a polished Si substrate. The nominal thicknesses of Si and Mo layers were 25.5 Å and 8.9 Å, respectively, giving a bilayer period of 34.4 Å in excellent agreement with the 34.4 Å as determined by Cu K$\alpha$ reflectivity at 8.04 keV. Hard X-ray photoelectron spectroscopy measurements were performed at beamline 9.3.1 of the Advanced Light Source at Lawrence Berkeley National Laboratory, with 3100 eV photon energy.

The problem definition provided to the optimizer is given in Table 2 and in Figure 5a. There are 5 layer thicknesses and 4 interdiffusion lengths to be optimized: a total of 9 parameters. The repeating part (80 repeats) is defined with $d_{Mo}$ (repeating Mo thickness), $d_{Si,ML}$ (repeating Si thickness), and $r_{ML}$ (repeating interdiffusion length between Mo and Si). The non-repeating part consists of $d_{Si,top}$ (thickness of the topmost Si layer), $d_{SiO2}$ (thickness of the native oxide — SiO$_2$), and $d_C$ (thickness of the surface carbon contamination), as well as the corresponding interdiffusion lengths/roughnesses for the top three layers: $r_{Si,top}$, $r_{SiO2}$, and $r_C$, respectively. The interdiffusion length between the topmost Si layer and the topmost Mo layer is also defined to be equal to $r_{ML}$ (i.e. the repeating interdiffusion length). The RCs were normalized by dividing each point by the second-order polynomial background that was fitted to the off-Bragg tails of the RCs.

Table 2: Limits and step sizes of the parameters in Example 2

|  | min [Å] | step [Å] | max [Å] |
|---|---|---|---|
| $d_C$ | 0 | 0.2 | 10 |
| $d_{SiO2}$ | 0 | 0.2 | 50 |
| $d_{Si,top}$ | 0 | 0.2 | 30 |
| $d_{Mo}$ | 6 | 0.2 | 10 |
| $d_{Si,ML}$ | 24 | 0.2 | 30 |
| $r_C$ | 0 | 0.5 | 8 |
| $r_{SiO2}$ | 0 | 0.5 | 8 |
| $r_{Si,top}$ | 0 | 0.5 | 8 |
| $r_{ML}$ | 0 | 0.5 | 6 |

SWOPT and the random search were run 10 times with 10 different random initial guesses and 10,000 YXRO simulations each time. Figure 5b shows the RCs corresponding to the 5$^{th}$ best, or approximately median solution of SWOPT, together with the experimental RCs. Figure 5c-d-e show the evolution of the best R value found so far as a function of the number of simulations performed. While Figure 5d and Figure 5e show the progress of the 10 individual runs for SWOPT and the random search, Figure 5c compares the averages of these two approaches — standard deviations are included in lighter colors. The average curves in Figure 5c show that SWOPT needed about an order of magnitude fewer evaluations than the random search to reach a certain R value. The standard deviation for SWOPT was also smaller.

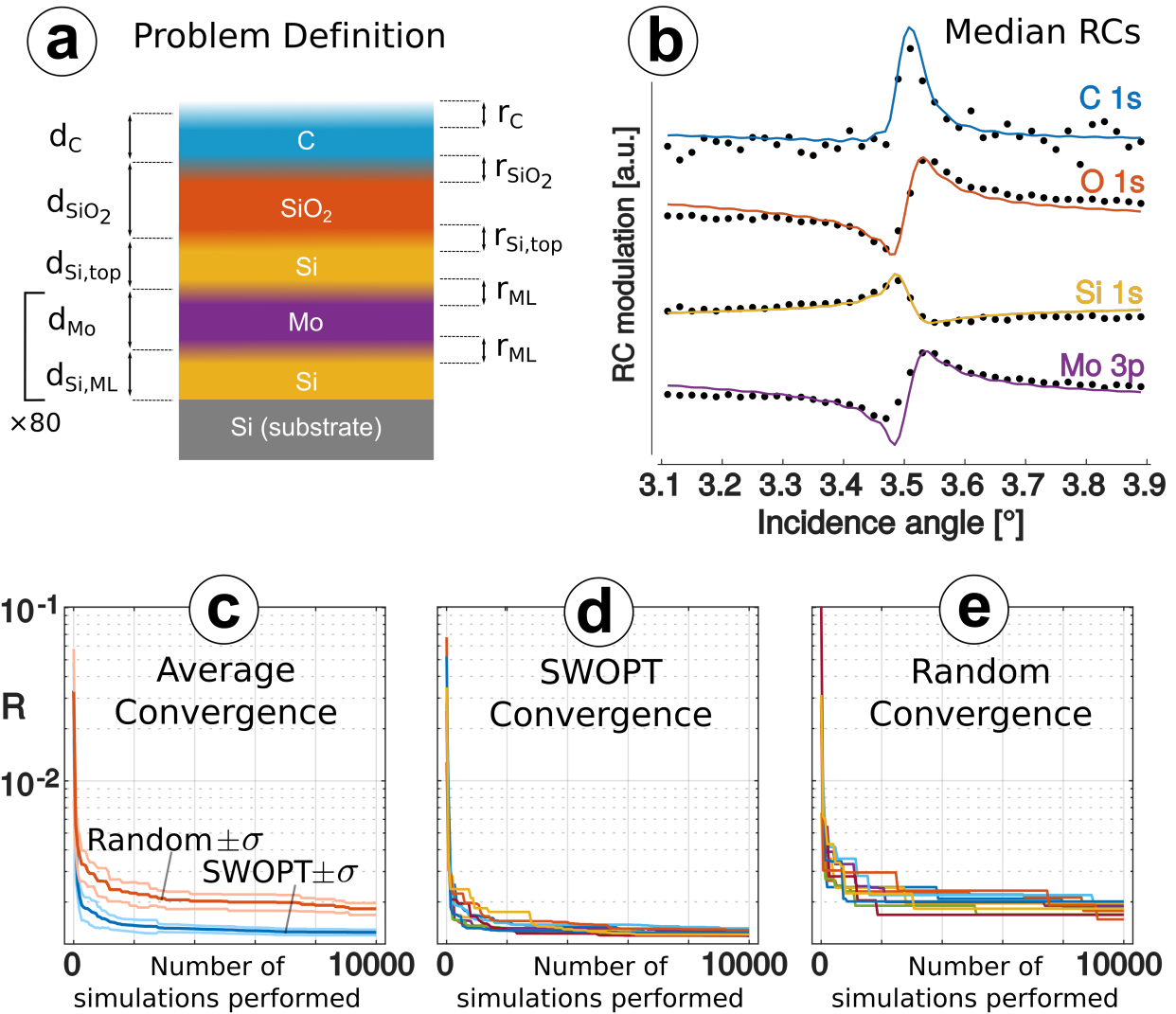

Figure 5: (a) Problem definition for Example 2; (b) experimental (solid circles) and median SWOPT-found (curves) RCs for C 1s, O 1s, Si 1s (elemental), and Mo 3p; best R value found as a function of the number of simulations performed for 10 different runs: (c) averages with standard deviations, (d) SWOPT for individual runs, (e) random search for individual runs.

The median (5th best), mean — with standard deviations — and minimum/maximum solutions for each parameter are given in Table 3 for both SWOPT and the random search. The two algorithms give similar results for layer thicknesses. The results of SWOPT have smaller standard deviations, which is likely an indication of better convergence. The carbon layer thickness "$d_C$" has a large relative standard deviation for SWOPT (71%) probably because it's a small fraction (~10%) of the standing-wave period and the RCs are normalized (i.e. absolute intensity information is lost). Note that the effective period of the multilayer is 25.2 + 8.2 = 33.4 Å, within about 3% of the 34.4 Å determined by hard X-ray diffraction.

An Efficient Algorithm for Automatic Structure Optimization
in X-ray Standing-Wave ExperimentsTable 3: Layer thicknesses obtained with SWOPT and the random search algorithm for the real Si/Mo multilayer case. "Median" refers to the thicknesses obtained in the 5[th] best solution of 10 trials with SWOPT and the random search ("Random"). "min/max" refers to the minimum/maximum value found for the parameter within the 10 runs. Mean values are reported for completeness; they are ensemble averages and don't necessarily correspond to a solution of SWOPT or random search.

|  | median [Å] | | mean (st.dev.) [Å] | | min. [Å] / max. [Å] | |
| --- | --- | --- | --- | --- | --- | --- |
|  | SWOPT | Random | SWOPT | Random | SWOPT | Random |
| $d_C$ | 6.8 | 0.0 | 4.5 (3.2) | 3.2 (2.1) | 0.6 / 9.8 | 0.0 / 5.8 |
| $d_{SiO2}$ | 10.0 | 11.0 | 11.2 (1.6) | 11.1 (2.4) | 8.0 / 13.2 | 7.8 / 15.0 |
| $d_{Si,top}$ | 25.0 | 25.6 | 25.1 (0.2) | 25.3 (0.8) | 24.8 / 25.4 | 24.0 / 26.6 |
| $d_{Mo}$ | 8.2 | 9.2 | 8.9 (0.8) | 8.6 (0.7) | 8.2 / 10.0 | 7.4 / 9.6 |
| $d_{Si,ML}$ | 25.2 | 25.4 | 24.9 (0.3) | 25.2 (0.6) | 24.6 / 25.4 | 24.0 / 26.2 |
| $r_C$ | 3.5 | 2.0 | 5.0 (1.6) | 3.2 (2.1) | 3.5 / 7.5 | 0.5 / 7.0 |
| $r_{SiO2}$ | 8.0 | 1.0 | 8.0 (0.0) | 3.5 (2.6) | 8.0 / 8.0 | 0.0 / 8.0 |
| $r_{Si,top}$ | 0.5 | 1.0 | 0.4 (0.8) | 3.0 (1.6) | 0.0 / 2.5 | 1.0 / 5.5 |
| $r_{ML}$ | 4.0 | 5.0 | 4.5 (0.4) | 4.3 (0.6) | 4.0 / 5.0 | 3.0 / 5.0 |

### 3.3.   Example 3: Cs+Na Aqueous Solution/$Fe_2O_3$/Si-Mo Multilayer Mirror

The third example is the re-evaluation of data that was published earlier [13], with all details discussed in this reference. In addition to the random search method, we compare the performance of SWOPT to the manual analysis performed by Nemšák et al. The original analysis of this problem took about two weeks with frequent human computer input. With SWOPT, it took the user ~1 h to set up the program definition, which is rather advanced compared to examples 1 and 2. After this, a single optimization run with 10,000 evaluations finished within ~6 h with no further human interaction required. This is thus roughly a factor 50 reduction in analysis time. Considering that 10,000 evaluations is at least twice of what is needed to obtain a solution akin to Nemšák's, we can say that a factor of ~100 reduction in data analysis time has been achieved for this particular example; and of course, human bias in varying parameters has also been removed.

In more detail, the sample is a $Fe_2O_3$ layer deposited on a Si/Mo MLM that is identical to the one in the previous example. A thin NaOH+CsOH layer was dropcast on $Fe_2O_3$ from an aqueous solution. The sample was hydrated *in situ* with 0.4 Torr $H_2O$ at 2.5 °C, which corresponds to

~8% relative humidity. The ions are thus expected to be mobile. This requires a more complex problem definition than the other examples, whereby the Na and Cs layers (i.e. the volumes that contain Na$^+$ and Cs$^+$ ions) can overlap and are free to move within the liquid layer. The parameterization of SWOPT, which includes the full functionality of YXRO-Ultra scripts, permits defining this complex problem in a complete way directly amenable to the BBO.

Figure 6a shows the problem definition where $d_{Fe2O3}$, $d_{H2O}$, $d_{Na}$, and $d_{Cs}$ are the thicknesses of the corresponding layers. To model the mobility of the Cs and Na layers, we introduced a layer above ($b_{Cs,top}$ and $b_{Na,top}$) and a layer below ($b_{Cs,bot}$ and $b_{Na,bot}$) each, whose thicknesses can vary independently. One condition, for this particular problem, is that:

$$d_{H_2O} = b_{Na,top} + d_{Na} + b_{Na,bot} = b_{Cs,top} + d_{Cs} + b_{Cs,bot},$$

that is, both ions must be found somewhere in the water layer of thickness $d_{H2O}$.

The SWOPT and the random search were run 10 times with 10 different initial random guesses and 10,000 YXRO simulations each time. Figures 6b and 6c show the RCs for the median (5$^{th}$ best) solution of SWOPT, and the solution of the manual analysis, respectively. Figure 6d and 6f show the convergence of individual runs for SWOPT and the random search, respectively. Figure 6d compares the averages of these two approaches — standard deviations are included in lighter colors. The average curves show that SWOPT outperforms random search — within 1 standard deviation — after ~500 iterations. The relatively small standard deviation for SWOPT indicates that it likely convergences close to the global minimum.

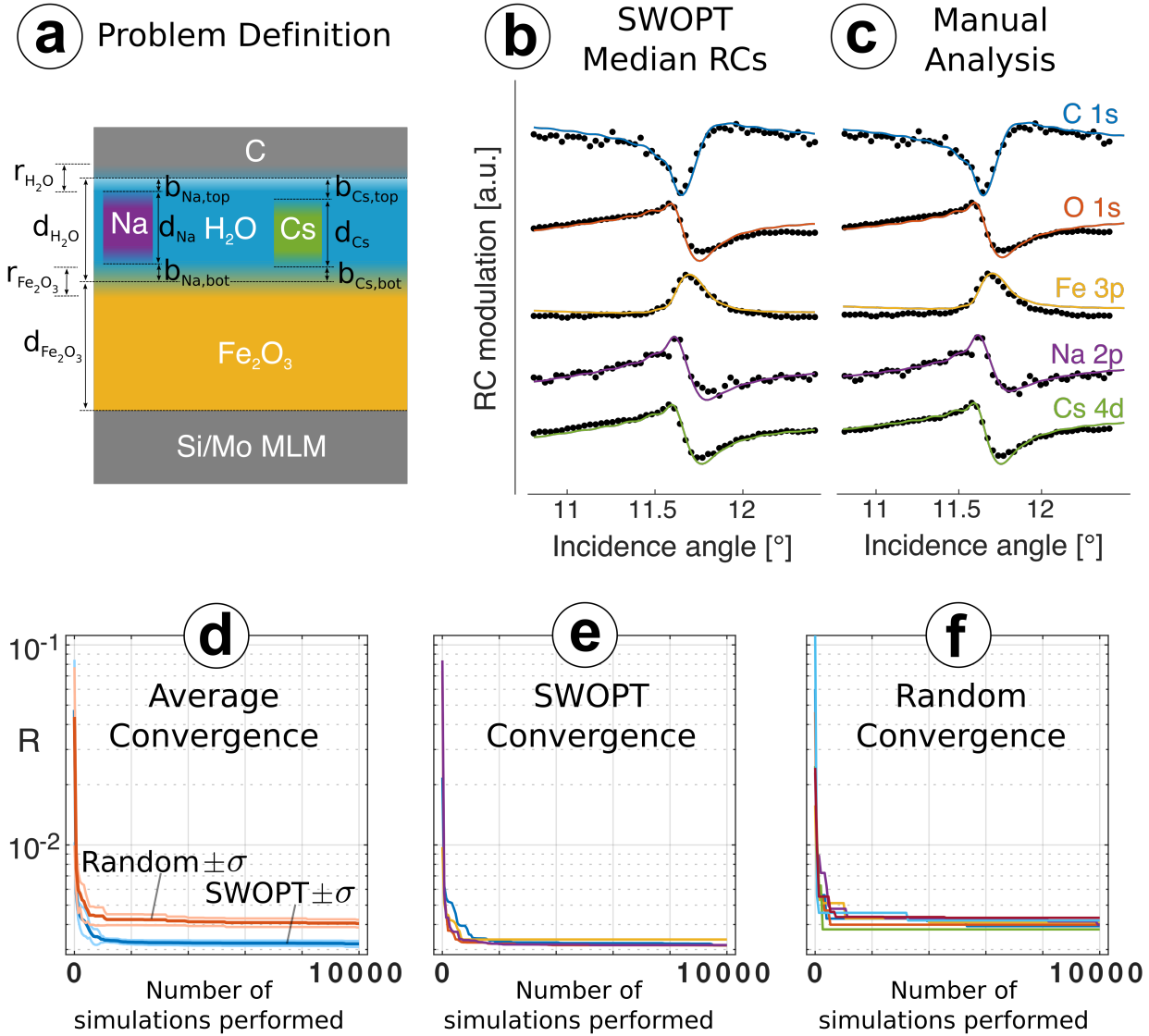

Figure 6: (a) Problem definition for Example 3; (b) solutions for the problem obtained via SWOPT, for the 5$^{th}$-best ≈ median solution showing experimental (solid circles) and computed (curves) RCs for C 1s (aliphatic), O 1s (everything except $Fe_2O_3$ and $H_2O(g)$), Fe 3p, Na 2p, and Cs 4d; (c) same as (b) but the solution was found manually by Nemšák et al. [13]; best R value found as a function of the number of simulations performed for 10 different runs: (d) averages with standard deviations, (e) SWOPT, (f) random.

The results extracted from the median SWOPT solution agree — as presented in Table 4 — extremely well with the parameters optimized by the much less sophisticated manual optimization, which essentially is a brute force method that optimized three parameters from the set at a time. Figure 6g presents the detailed distributions derived for this sample from SWOPT. The deviations between the two found solutions are within a few Å, which is also a limiting precision of this type of experiment, considering that the SW period is roughly one order-of-magnitude larger than these deviations. The solutions of all 10 trials of SWOPT optimizations unanimously confirm the conclusion that Na$^+$ ions are most likely stripped of their solvation shells and are adsorbed specifically at the liquid/hematite interface. For the Cs$^+$ ions, not every trial resulted in a sizable spacing between hematite and liquid, which is represented by $d_{Cs,bot}$ in Table 4 with a mean value of $d_{Cs,bot}$ = 0.7 Å ± 0.6 Å. This variation is explained by the fact that the exact shape of the RC, and therefore also the objective function value, are much less dependent on the bottom interfaces of the liquid and ion distributions compared to the top interfaces. However, judging by the best of the 10 SWOPT solutions we found a spacing of 1.3 Å between the hematite and the Cs$^+$ ions, which is consistent with the results by Nemšák et al. [13]. These results are furthermore consistent with earlier studies of alkali adsorption on hematite [26, 27] as well as different behavior between Na$^+$ and Cs$^+$ at the liquid/vapor interface [28, 29]. SWOPT results, just like the analysis by Nemšák et al., indicate that Cs$^+$ ions are directly at the liquid/vapor interface — probably with a partial solvation shell — and Na$^+$ ions are excluded from this interface to a depth of ~4 Å.

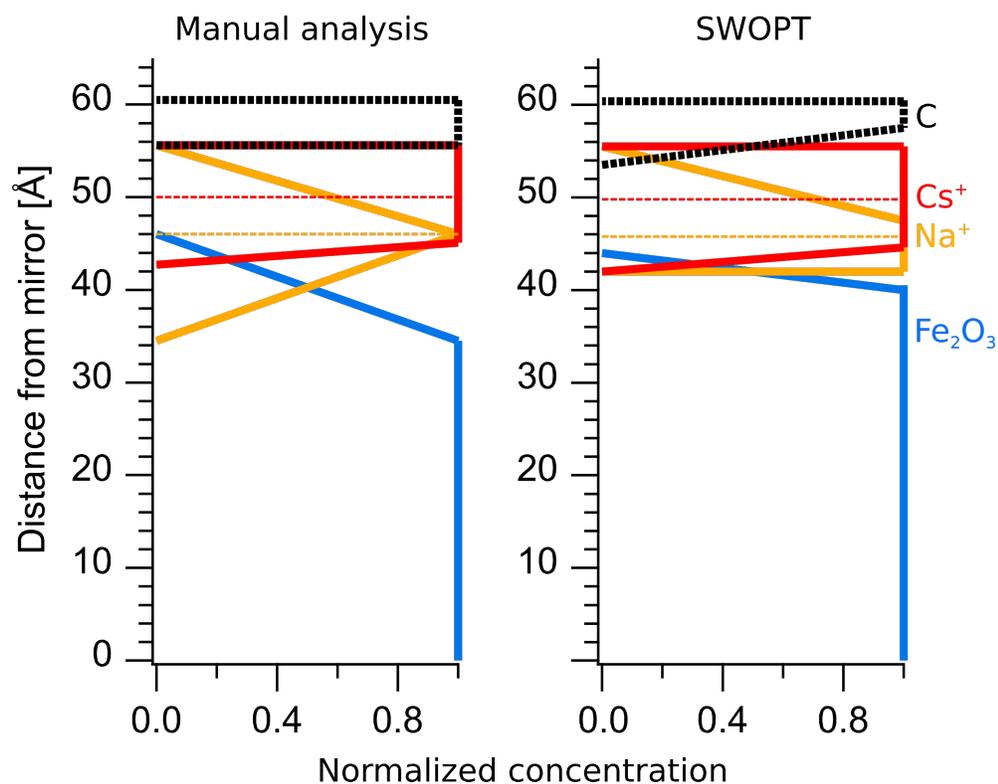

Figure 7: Comparison of element distribution determined by manual analysis [13] to the best result of the 10 SWOPT trails that are presented here. (Note that the original figure in ref. [13] showed interdiffusion lengths that were too low by a factor of 2, which was corrected here.)

Figure 7 shows the element distributions that were determined by the manual analysis of Nemšák et al. compared to the best of the 10 SWOPT trials. Despite slight differences in the exact shape of the distributions, the overall agreement is good. Especially the main conclusions of the original study — the repulsion of Cs$^+$ from the hematite, and of Na$^+$ from the liquid/vapor interface — are reproduced by SWOPT, even if the program had to be run several times at 10000 iterations to find the best result. The best, worst, and median (5[th] best) objective function values that we received using SWOPT were R = 0.00314, 0.00337, and 0.00316, respectively, compared to R = 0.0039 in the manual analysis. Therefore, we can conclude that SWOPT is at least as reliable as the manual analysis. In addition, it is much less dependent on the intuition of the scientist analyzing the data. While the results are similar, the manual analysis relied largely on the experience of the scientist to find a good initial guess of the structure, and then was limited to varying a maximum of three parameters simultaneously to optimize the sample properties. This not only explains the significant amount of human interaction needed in the manual analysis, but it also makes it unlikely that a scientist with less experience in analyzing SW-PES data could find a sufficiently accurate solution, at least not without expending a much greater amount of time. Our new program SWOPT does not rely on initial and often subjective guesses beyond the upper and lower limit for each optimization parameter.

Table 4: Layer thickness and roughness values obtained manually [13], from the SWOPT program, and the random search, for the SWAPPS example.

| | manual analysis [Å] | median [Å] | | mean (st.dev.) [Å] | | min [Å] / max [Å] | |
|---|---|---|---|---|---|---|---|
| | | SWOPT | Random | SWOPT | Random | SWOPT | Random |
| $d_{H2O}$ | 15.1 | 13 | 14 | 13.1 (0.5) | 13.8 (0.5) | 12.5 / 13.5 | 13.0 / 14.5 |
| $d_{Na,top}$ | 4.2 | 4.6 | 3.8 | 4.7 (0.9) | 4.0 (1.1) | 4.1 / 6.3 | 2.6 / 5.8 |
| $d_{Na}$ | 10.9 | 8.5 | 9.8 | 8.4 (1.2) | 9.0 (1.5) | 6.3 / 9.5 | 7.0 / 11.6 |
| $d_{Na,bot}$ | 0.0 | 0.0 | 0.4 | 0.0 (0.0) | 0.7 (0.6) | 0.0 / 0.0 | 0.1 / 1.9 |
| $d_{Cs,top}$ | 0.0 | 0.0 | 0.0 | 0.0 (0.0) | 0.3 (0.4) | 0.0 / 0.1 | 0.0 / 1.4 |
| $d_{Cs}$ | 11.7 | 12.4 | 14 | 12.4 (0.3) | 12.4 (1.7) | 12.2 / 13.0 | 8.5 / 14 |
| $d_{Cs,bot}$ | 3.4 | 0.6 | 0.0 | 0.7 (0.6) | 1.0 (1.1) | 0.0 / 1.3 | 0.0 / 3.2 |
| $d_{Fe2O3}$ | 40.5 | 42.5 | 42.0 | 42.4 (0.4) | 41.9 (0.3) | 42.0 / 43.0 | 41.5 / 52.5 |
| $r_{H2O}$ | 0.0 | 2.0 | 3.0 | 2.0 (0.0) | 2.5 (0.5) | 2.0 / 2.0 | 2.0 / 3.5 |
| $r_{Fe2O3}$ | 3.0 | 2.0 | 2.0 | 2.0 (0.0) | 3.1 (0.8) | 2.0 / 2.0 | 2.0 / 4.5 |

# 4. Conclusions

We show that our new computer program (SWOPT) solves the structure optimization problem in SW-PES problems involving multilayered samples quickly (minutes to hours), and without extensive human interaction. We compared the performance of our black-box optimizer (BBO) algorithm to a random search method by running each 10 times for three test cases. In every case the SWOPT significantly outperformed the random search.

The complex hypothetical test case involving "pseudo-experimental" data from 5 layers on a Mo/Si multilayer showed that for the layers closer to the surface, thicknesses could be analyzed more accurately. The relative standard deviations of layer thicknesses for 10 independent solutions were in the range of 2–8 %.

The experimental test case for a bare Si/Mo multilayer, which is a relatively simple problem, showed that the best of the 10 random searches gave similar results to SWOPT after 10,000 iterations. However, SWOPT not only found the solution more efficiently, it also consistently found a good solution in all 10 trials. Thus, SWOPT can be relied on for finding a solution under time limitations or with little computational resources and proves to be far more reliable after only a single trial run.

We also used SWOPT to reanalyze the results of an earlier SWAPPS study that investigated the $Cs^+/Na^+$ distribution normal to the surface in a liquid layer on $Fe_2O_3$. SWOPT reproduced the results found earlier within ~1 Å for all but one parameter: only an ~3 Å difference with the manual analysis was found for the $Cs^+/Fe_2O_3$ gap, which is within the precision of the experiment.

The versatility of SWOPT is an important advantage for tackling future SW-PES problems with greater complexity. The ~10 fold speed gained through "YXRO-Ultra", and the extra speed through the BBO algorithm, which ultimately leads to analyses about 50-100x faster, makes this program an invaluable tool for "realtime" analysis of data during synchrotron experiments. We will make SWOPT available online to the community in the future.

# 5. Acknowledgments

We acknowledge support by the Director, Office of Science, Office of Basic Energy Sciences, and by the Division of Chemical Sciences, Geosciences and Biosciences of the US Department of Energy at LBNL under Contract No. DE-AC02-05CH11231, and salary support for M.G and C.S.F from a grant from the Basic Energy Sciences Division of Materials Physics and Engineering at UC Davis, No. DE-SC0014697. This material is also based upon work supported by the U.S. Department of Energy, Office of Science, Office of Advanced Scientific Computing Research, Applied Mathematics program under contract number DEAC02005CH11231. We also acknowledge the support of CAMERA—the Center for Advanced Mathematics for Energy Research Applications at Lawrence Berkeley National Laboratory, and the Air Quality Research Center and the Physics Department, University of California Davis, Davis, CA USA, who maintain and operate the computer cluster BEAR, on which the program YXRO Ultra was tested, developed, and used.

# 6. Appendix – Surrogate Optimization Algorithm For Constrained Integer Problems

Below, we describe the detailed individual steps of the black-box optimization method (BBO).

| | |
|---|---|
| Step 1: | Create an initial experimental design with $\min\{K, d+1\}$ points, where $K$ is the total number of possible solutions (including infeasible solutions). We use a symmetric Latin hypercube design and round the parameter values to the closest integers. Denote the solution set by $\mathcal{S}_0$. |
| Step 2: | Compute the (computationally cheap) constraint function values and discard the infeasible points. Denote the remaining solution set by $\widetilde{\mathcal{S}_0}$. |
| Step 3: | If the rank condition for the matrix $P$ (see Eq. 9) corresponding to the points in $\widetilde{\mathcal{S}_0}$ is not satisfied, keep adding feasible points to $\widetilde{\mathcal{S}_0}$ until the rank condition is satisfied. Denote the final set of points by $\mathcal{S}$. |
| Step 4: | Compute the objective function values at all points in $\mathcal{S}$, denote the vector of function values by $F$. |
| Step 5: | Find the point in $\mathcal{S}$ with the smallest objective function value. Denote the point by $x_{\text{best}}$ and its function value by $f_{\text{best}}$. |
| Step 6: | Iterate while stopping criterion is not met: |
| Step 6a: | Given the points in $\mathcal{S}$ and their function values $F$, compute the parameters of the RBF surrogate model, $s(x)$ (see Eq. 8). |
| Step 6b: | Create a large set of candidate points around $x_{\text{best}}$ by perturbing the variable values of $x_{\text{best}}$ and by uniformly selecting points from the whole variable domain. All created candidate points satisfy the integrality constraints. |
| Step 6c: | Discard all candidate points that do not satisfy the constraints $c_j$. Denote the remaining candidate points by $\chi_1, \dots, \chi_M$. |
| Step 6d: | Use the surrogate model to predict the objective function values for all candidate points and denote the set of predicted values by $s(\chi_1), \dots, s(\chi_M)\}$ and scale these values to [0,1] such that the smallest predicted value is set to 0 and the largest value is set to 1. Denote the set of scaled values by $V_{\mathcal{R}}$. |
| Step 6e: | Compute the distance of all candidate points to the set $\mathcal{S}$, denote the set by $\{\Delta(\chi_1, \mathcal{S}), \dots, \Delta(\chi_M, \mathcal{S})\}$. Scale the distance values to [0,1] such that the candidate with the largest distance obtains score 0 and the candidate |
|  | closest to $\mathcal{S}$ obtains score 1. Denote the set of scaled distance values by $V_\mathcal{D}$. |
| Step 6f: | Compute a weighted sum of both scores $\mathcal{V} = \omega V_\mathcal{R} + (1-\omega)V_\mathcal{D}$ for each candidate point and select the candidate with the lowest score as new evaluation point, $x_{\text{new}}$. |
| Step 6g: | Compute the expensive objective function at the new point, $f(x_{\text{new}})$. |
| Step 6h: | If $f(x_{\text{new}}) < f_{\text{best}}$ |
|  | $f_{\text{best}} = f(x_{\text{new}})$ |
|  | $x_{\text{best}} = x_{\text{new}}$ |
| Step 6i: | $\mathcal{S} = \mathcal{S} \cup \{x_{\text{new}}\}; F = F \cup \{f(x_{\text{new}})\}$. |
| Step 7: | Return the best solution found. |

In Step 1 of the algorithm, if $K$ is small enough, we may do a complete enumeration of all possible solutions, discard all infeasible points, and only sample from the remaining set of points. This approach makes the creation of the initial sample points and the candidate points in Step 6b easier. In the iterative sampling procedure, we ensure that no point will be selected for evaluation more than once. Since the objective function is deterministic, repeatedly evaluating at the same point will not add new information.

In Step 6d, we scale the predicted objective function values of all candidates to [0,1] according to

$$V_\mathcal{R}(\chi_l) = \frac{s(\chi_l) - s_{\min}}{s_{\max} - s_{\min}}, l = 1, \ldots, M, s_{\min} = \min\{s(\chi_l), l = 1, \ldots, M\}, s_{\max} = \max\{s(\chi_l), l = 1, \ldots, M\}.$$

The distance scores in Step 6e are computed as follows:

$$\Delta(\chi_l) = \min_{x_k \in \mathcal{S}} \|\chi_l - x_k\|_2, l = 1, \ldots, M.$$

The [0,1] scaling is done according to

$$V_\mathcal{D}(\chi_l) = \frac{\Delta_{\max} - \Delta(\chi_l)}{\Delta_{\max} - \Delta_{\min}}, l = 1, \ldots, M, \Delta_{\max} = \max\{\Delta_l, l = 1, \ldots, M\}, \Delta_{\min} = \min\{\Delta_l, l = 1, \ldots, M\}$$

(see also [30]). In Step 6f, we compute a weighted score between the surrogate model prediction and the distance to already evaluated points:

$$\mathcal{V}(\chi_l) = \omega V_\mathcal{R}(\chi_l) + (1-\omega)V_\mathcal{D}(\chi_l), l = 1, \ldots, M$$

The weights $\omega$ are selected from a range of possibilities, namely $\mathcal{W} = \langle 1, 0.9, 0.75, 0.6, 0.5, 0.35, 0.25, 0 \rangle$. When selecting only a single point in each iteration for doing the expensive function evaluation, we cycle through this weight pattern. Our implementation allows also to select more than one point in each iteration, in which case we select 8 points (the length of the weight pattern $\mathcal{W}$) and each point is chosen based on using a different weight. We update the

...............1

best point and the best function value found so far as well as the set of already evaluated points and the corresponding set of function values. The algorithm stops once we have reached a predefined maximum number of objective function evaluations.

The convergence of the algorithm to the global optimum follows from a simple counting argument. As there exist only a finite number of possible solutions and the algorithm never evaluates a parameter vector more than once, it follows that as the number of evaluations approaches $K$, we will eventually sample at the global optimum. However, in practice, we cannot allow $K$ evaluations and we have to be satisfied with the best solution we can obtain within our computation time budget. Typically we let the program run through about 10,000 iterations.

An Efficient Algorithm for Automatic Structure Optimization
in X-ray Standing-Wave Experiments[13] S. Nemšák, et al., Concentration and chemical-state profiles at heterogeneous interfaces with sub-nm accuracy from standing-wave ambient-pressure photoemission, Nat Commun, 5 (2014) http://dx.doi.org/10.1038/ncomms6441.

[14] O. Karslıoğlu, et al., Aqueous solution/metal interfaces investigated in operando by photoelectron spectroscopy, Faraday Discuss, 180 (2015) 35-53, http://dx.doi.org/10.1039/c5fd00003c.

[15] S. Nemšák, et al., Element- and momentum- resolved electronic structure of the dilute magnetic semiconductor Ga1−xMnxAs, arXiv e-print, (2018).

[16] S.H. Yang, et al., Making use of x-ray optical effects in photoelectron-, Auger electron-, and x-ray emission spectroscopies: Total reflection, standing-wave excitation, and resonant effects, J Appl Phys, 113 (2013) http://dx.doi.org/10.1063/1.4790171.

[17] J.B. Pendry, Reliability Factors for Leed Calculations, J Phys C Solid State, 13 (1980) 937-944, http://dx.doi.org/10.1088/0022-3719/13/5/024.

[18] S.P. Tear, et al., A Comparison of Reliability (R) Factors in a Leed Structural-Analysis of the Copper (111) Surface, J Phys C Solid State, 14 (1981) 3297-3311, http://dx.doi.org/10.1088/0022-3719/14/22/023.

[19] A.J. Booker, et al., A rigorous framework for optimization of expensive functions by surrogates, Struct Optimization, 17 (1999) 1-13, http://dx.doi.org/10.1007/BF01197708.

[20] M.J.D. Powell, Recent research at Cambridge on radial basis functions, in, Birkhäuser Basel, Basel, 1999, pp. 215-232.

[21] G. Matheron, Principles of geostatistics, Economic Geology, 58 (1963) 1246-1266, http://dx.doi.org/10.2113/gsecongeo.58.8.1246.

[22] J.H. Friedman, Multivariate Adaptive Regression Splines, Ann Stat, 19 (1991) 1-67, http://dx.doi.org/10.1214/aos/1176347963.

[23] R.H. Myers, D.C. Montgomery, Response Surface Methodology: Process and Product Optimization Using Designed Experiments, John Wiley & Sons, Inc., 1995.

[24] J. Müller, R. Piché, Mixture surrogate models based on Dempster-Shafer theory for global optimization problems, J Global Optim, 51 (2011) 79-104, http://dx.doi.org/10.1007/s10898-010-9620-y.

[25] J. Müller, et al., SO-I: a surrogate model algorithm for expensive nonlinear integer programming problems including global optimization applications, J Global Optim, 59 (2014) 865-889, http://dx.doi.org/10.1007/s10898-013-0101-y.

[26] H. Amhamdi, et al., Effect of urea on the stability of ferric oxide hydrosols, Colloid Surface A, 125 (1997) 1-3, http://dx.doi.org/10.1016/S0927-7757(96)03880-0.

[27] L.J. Kirwan, et al., An in situ FTIR-ATR study of polyacrylate adsorbed onto hematite at high pH and high ionic strength, Langmuir, 20 (2004) 4093-4100, http://dx.doi.org/10.1021/la036248u.

[28] W. Bu, et al., Ion distributions at charged aqueous surfaces by near-resonance X-ray spectroscopy, J Synchrotron Radiat, 13 (2006) 459-463, http://dx.doi.org/10.1107/S0909049506038635.

[29] W. Hua, et al., Cation Effects on Interfacial Water Organization of Aqueous Chloride Solutions. I. Monovalent Cations: Li+, Na+, K+, and NH4+, J Phys Chem B, 118 (2014) 8433-8440, http://dx.doi.org/10.1021/jp503132m.

[30] R.G. Regis, C.A. Shoemaker, A stochastic radial basis function method for the global optimization of expensive functions, Informs J Comput, 19 (2007) 497-509, http://dx.doi.org/10.1287/ijoc.1060.0182.25